\documentclass[journal,10pt,final]{IEEEtran}

 \usepackage{graphicx}
 \graphicspath{{/home/s0897243/Downloads/oldtran/}{/home/s0897243/Downloads/oldtran/}}
 \graphicspath{{figure/}}
 \usepackage{float}
 \usepackage{amsmath}
 \usepackage{amssymb}
 \usepackage{mathrsfs}
 \usepackage{mathtext}
 \usepackage{url}

\begin{document}

\title{Powerline Communications Channel Modelling Methodology Based on Statistical Features}

\author{Bo~Tan,~\IEEEmembership{Student Member,~IEEE,}
        and~John~S.~Thompson,~\IEEEmembership{Member,~IEEE}% <-this % stops a space
\thanks{B. Tan and J.S. Thompson is with Joint Research Institute for Signal and Image Processing, Department of Electronic Engineering, University of Edinburgh, Edinburgh, EH9 3JL, UK. (Email: {B.Tan, John.Thompson}@ed.ac.uk)}}

\maketitle

\begin{abstract}
This paper proposes a new channel modelling method for powerline communications networks based on the multipath profile in the time domain. The new channel model is developed to be applied in a range of Powerline Communications (PLC) research topics such as impulse noise modelling, deployment and coverage studies, and communications theory analysis. To develop the methodology, channels are categorised according to their propagation distance and power delay profile. The statistical multipath parameters such as path arrival time, magnitude and interval for each category are analyzed to build the model. Each generated channel based on the proposed statistical model represents a different realisation of a PLC network. Simulation results in similar the time and frequency domains show that the proposed statistical modelling method, which integrates the impact of network topology presents the PLC channel features as the underlying transmission line theory model. Furthermore, two potential application scenarios are described to show the channel model is applicable to capacity analysis and correlated impulse noise modelling for PLC networks. 
\end{abstract}


\IEEEpeerreviewmaketitle

% \newpage 
%%%%%%%%%%%%%%%%%%%%%%%%%%%%%%%%%%%%%%%%%%%%%%%%%%%%%%%%%%%%%%%%%%%%%%%%%%%%%%%%%%%%%%%%%%%%
%		Introduction
%%%%%%%%%%%%%%%%%%%%%%%%%%%%%%%%%%%%%%%%%%%%%%%%%%%%%%%%%%%%%%%%%%%%%%%%%%%%%%%%%%%%%%%%%%%%
\section{Introduction}
With the spread of Smart Grid concepts, powerline communications which has been used for voice transmission technology since 1920s \cite{History_PLC} has become one important option for the required data communications infrastructure. It has become an attractive alternative for conventional wired and wireless in-door data transmission due to the development of robust modulation, channel coding, and digital signal processing technologies. Driven by the increasing market demand, a serial of standards such as IEEE P1901/1901.2, ITU-T G.hn/G.cx/G.hnem have been released to meet the requirements of in-home broadband communications, Smart Grid applications, and in-home energy management \cite{PLC_Standards}. However, there is lack of a common agreement on the most suitable channel model. Thus, it is important to develop accurate channel models to evaluate the above technologies.

The currently proposed PLC channel modeling approaches in the literature can be classified into two main categories: the multipath approach \cite{Multi_path_Zimmermann}-\cite{Multi_path_Anastasiadou} and the transmission line theory approach \cite{MTL_Galli}, \cite{MTL_Meng}. For the multipath model, it models the reflection signals caused by impedance mismatch on the branches of a PLC network such as the segment $T_{B}$-$T_{D}$ in Fig.~\ref{Fig:T_type_top}. The key issue of this model is to select accurately the parameters of the signal propagation properties and the multipath effect, which is based on the measurement of actual channel transfer characteristics. Otherwise, it is difficult to evaluate the propagation properties of each path individually among the potentially infinite series of reflection signals. In particular, when the network scale is large and complex, to consider all the propagation paths and corresponding branch locations, path lengths and impedance mismatches becomes much too complex. For the transmission line theory model, the PLC network is considered as a cascade of two-port network segments. By using transmission matrices for each segment, the impact of each individual path in the time domain is automatically incorporated into the frequency domain transfer function. In addition, 
transmission line theory splits the network into several independent components, thus, this modeling approach can be applied to varying network topologies. However, it is still a time-consuming task to use transmission line theory to model the PLC channel and difficult to apply it in practice when the topology is complex. There are also some extended modeling methods based on the multipath and transmission line theory approaches as shown in \cite{MTL_Galli_Deter} and \cite{Mulit_path_Xin}. In \cite{MTL_Galli_Deter}, the authors proposed a deterministic frequency domain model to standardize the process of modeling PLC transfer functions. In \cite{Mulit_path_Xin} Ding brought the multipath model into a more complex topology scenario and integrated the iterative path property calculations into the metrics. However, the modeling methods are still limited by the network topology. Due to the drawbacks of the above models, statistical models are proposed in \cite{Statistic_Galli}  -\cite{Statistic_Andrea_II}, which integrate the impact of the varying topology into statistical parameters, and also provide a platform for the PLC network deployment, coverage studies and communications theory analysis. These papers point out that the channel attenuation and Root-Mean-Square Delay-Spread (RMS-DS) of in-door PLC channels are correlated lognormal random variables. Reference \cite{Statistic_Galli} proposed a two-tap statistical model by setting the amplitude and spread delay for two channel taps following the Average Channel Gain (ACG) and the RMS-DS distribution in the paper. References \cite{Statistic_Andrea_I} and \cite{Statistic_Andrea_II} abstract the statistical channel parameters such as AGC, RMS-DS and path loss based on a randomly generated in-door PLC topology. The capacity of the PLC network is investigated according to the abstracted features. The statistical characteristics of the first arrival path is analyzed in detail in \cite{Statistical_firstpath}.

However, channel modelling is still an open research area, especially for time domain statistical models. The number of paths, the power delay profile, and the cable loss properties of the multipath PLC channel still need to be investigated in detail. It is important to develop a more precise channel statistical model for in-door PLC channels to reveal the path distribution,  path delay, path magnitude and phase which are vitally important for evaluating the potential of PLC. A new modeling method is proposed in this paper which uses analysis of the path magnitude, attenuation and path interval distribution and is verified against the modelling results from a transmission line theory simulation model. The key points of the new propagation model are follows:
\begin{enumerate}
 \item The arrival time and magnitude of the first arrival path in different channel are investigated in detail. The concept of a \textbf{Cluster} which indicates the transmission distance is introduced to identify the different arrival times of the first paths in different PLC deployments. In addition, the concept of a \textbf{Class} is developed to denote the degree of dispersion of a particular multipath channel. 
 \item The exponential Power Delay Profile (PDP) of the paths following the first path is derived according to the statistical data of the path magnitude. 
 \item The interval between adjacent paths for the multipath PLC channel is investigated. The General Extreme Value distribution is introduced to describe the path interval distribution in the PLC network..
 \item Two application scenarios are also introduced to show the efficiency of the proposed channel modelling method. The applications show that the model is suitable for communications theory analysis and for impulse noise modelling which is common noise source in PLC.
\end{enumerate}
The proposed statistical channel model is based on a group of network topologies described by a set of statistical parameters. Compared with other statistical PLC models, it provides the magnitude and delay for each individual path rather than just the AGC and RMS-DS. The impulse channel response for arbitrary Point to Point transmission distances within the scope can be derived from the proposed model.  
 
The paper is organized as follows. First, the multipath propagation behaviour and the basic time domain features are explained in Section II. Section III describes the random topology generator and the channel modeling procedure. The modeling results are verified in Section IV. Finally, Section V provides conclusions to the paper and describes future work. 

%%%%%%%%%%%%%%%%%%%%%%%%%%%%%%%%%%%%%%%%%%%%%%%%%%%%%%%%%%%%%%%%%%%%%%%%%%%%%%%%%%%%%%%%%%%%
% Time Domain Statistical Feature of Indoor PLC channel
%%%%%%%%%%%%%%%%%%%%%%%%%%%%%%%%%%%%%%%%%%%%%%%%%%%%%%%%%%%%%%%%%%%%%%%%%%%%%%%%%%%%%%%%%%%%
\section{Signal Propagation in PLC Network} \label{section: general_feature}
\subsection{Multipath Propagation in PLC Channel}
\begin{figure}
\begin{center}
 \includegraphics[scale=.4]{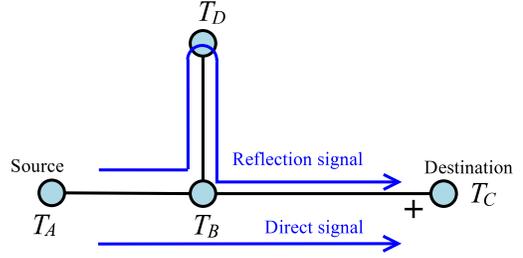}
 % Multipath_demo.eps: 128x0 pixel, 300dpi, 1.08x0.00 cm, bb=0 0 493 246
 \caption{Multipath propagation for a simple T type network}
 \label{Fig:T_type_top}
\end{center}
\end{figure}

It is a commonly agreed that an indoor PLC network often shows a tree type topology \cite{Book_Halid_treetop}. When a signal passes through the tree from source to destination, the signal energy will be split by the branches at the junctions, and reflected at the branch terminations due to impedance mismatches. Thus, the received signal at the destination can be considered as the combination of the signal propagating through the direct path from the transmitter and a group of reflections from network branches. Here a simple T type topology network, Fig.~\ref{Fig:T_type_top}, is used to demonstrate typical propagation behaviour.
The signal propagation paths for this type network can be written as: 
\linebreak
\linebreak
\vspace{0.5cm}
\begin{tabular}{ll}
Path 1 :& $T_{A}\rightarrow T_{B}\rightarrow T_{C}$ \\
Path 2 :& $T_{A}\rightarrow T_{B}\rightarrow T_{D}\rightarrow T_{B}\rightarrow T_{C}$ \\
Path 3 :& $T_{A}\rightarrow T_{B}\rightarrow T_{D}\rightarrow T_{B}\rightarrow T_{D}\rightarrow T_{B}\rightarrow T_{C}$ \\
 & $\cdots $  \\
Path \textit{N} :& $T_{A}\rightarrow T_{B}\rightarrow T_{D}\rightarrow \ldots T_{B}\rightarrow T_{C}$ \\
\end{tabular} 

Theoretically, the number of reflection paths can be infinite. Thus the channel impulse response for the $A\rightarrow C$ link can be expressed as: 
\begin{equation}\label{Equ:multipath_prototype}
 h\left(t \right) =\sum_{i=1}^{\infty} I_{i} \cdot \delta(t-T_{i})\cdot e^{j\varphi_{i}} 
\end{equation}
where, $I_{i}$, $T_{i}$ and $\varphi_{i}$ are the magnitude, delay and phase of the $i$th path respectively. 

For the parameters in the model of (\ref{Equ:multipath_prototype}), $I_{i}$ is impacted by the propagation constant and reflection coefficients of the junctions and branch terminations, which can be calculated accurately by using transmission line theory. When the network topology becomes complex, and the load impedances are random, it becomes very complex to evaluate these parameters for each path. Thus, the statistical features of the channel are evaluated in the later parts of this paper. The delay $T_{i}$ is a direct representation of the propagation distance of the $i$th path, $T_{i}=\frac{d_{i}}{v_{p}}$, where, $d_{i}$ is the propagation distance of $i$th path and $v_{p}$ is the signal propagation velocity in the power cable which, is determined by the dielectric constant of the insulating material. The scalar $\varphi_{i}$ represents the phase of the \textit{i}th path. In a multipath PLC channel, the first few paths will have a phase distribution determined by the propagation distance since they do not experience multiple reflections. For the following paths, the phases will follow a random distribution due to multipath reflections from multiple branches.
 
\subsection{A Practical In-door PLC Topology and Transfer Function} \label{Topology_MTL_model}
In order to capture the signal propagation behaviour characteristics of a large number of real PLC network, a randomly generated networks topology is used. In many paper such as \cite{Mulit_path_Xin}, \cite{Statistic_Andrea_I}and \cite{Book_Halid_treetop}, the tree type topology of the PLC network for both in-door and wide area network have been described. The PLC network can be considered as a series of branches connected by backbone cables. On each branch, there may be sub-branches spreading out to reach rooms in buildings or houses. Notice that all the sub-branches can be merged into the corresponding connected branch according to the impedance carry-back method in \cite{Statistic_Andrea_I}. Thus the PLC network can be considered equivalent to the topology shown in Fig.~\ref{Fig:Complex_top} which consists of backbone cables and first order branches. The components and configuration of the network topology are described as follows: 
\begin{figure}
\begin{center}
 \includegraphics[bb=0 0 645 328, scale=.3]{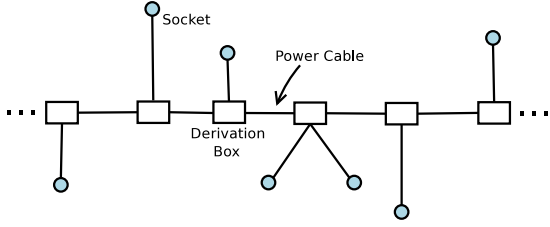}
 % Complex_demo.eps: -415946184x61 pixel, 300dpi, -3521677.50x0.52 cm, bb=0 0 645 328
 \caption{A typical network topology for the statistical model.}
 \label{Fig:Complex_top}
\end{center}
\end{figure}

\begin{itemize}
 \item \textbf{Components:} From Fig.~\ref{Fig:Complex_top} three components, namely cables, outlet (circles) and junctions (squares) are used to form a PLC network. Cables are used to connect the outlets and junctions. In this paper, the cable types NAYY35 and NAYY150, which are widely used for indoor power distribution are used for outlet-outlet and inter-junction connections respectively. The junctions can be a derivation box in practice. The outlets can be an open circuit power socket or a socket plugged with a appliance. Reflection signals occur at terminals with open sockets or mismatched appliances.

 \item \textbf{Branch density:} In powerline networks, the branch density may vary from scenario to scenario. In this paper, our focus is on the methodology for modelling the channel based on the statistical features of the channel impulse responses. A key issue relates to the ability to resolve multipath components in the time domain. In order to extract precise statistical properties from clearly resolved paths in the time domain, a low branch density which is identified as $\rho$ is used in this paper. As an example scenario, $\rho$ is set to 5 which indicates an average of 5 branches per 100 m of cable. The exact number of branches follows a Poisson distribution with mean 5 and this is used throughout this paper.

 \item \textbf{Branch location:} For a given transmission distance and number of branches, branches being uniformly distributed along the transmission path is a reasonable assumption which is also used in \cite{Statistic_Andrea_I}.

 \item \textbf{Branch length:} In \cite{Statistic_Andrea_I} three connection types are described. For each connection type, the Probability Density Function (PDF) of branch length is given as function of the side length of the cell in building. In this paper, the branch length is generated according the the PDFs in \cite{Statistic_Andrea_I} with a maximum side length of up to 20m.

 \item \textbf{Terminal load:} To approach a realistic scenario, half of the terminals are randomly set to open circuit. For the remaining sockets, the impedance are randomly allocated a discrete value between 5 ohms to 200 ohms with a 5 ohm interval. Similar assumptions can be also found in \cite{Mulit_path_Xin} and \cite{Statistic_Andrea_I}.
\end{itemize}

The voltage and current transfer characteristics of each segment in the above topology can be expressed by ABCD-parameters which can be illustrated by the Two-Port network (2PN) in Fig.~\ref{Fig:ABCD}:
\begin{figure}
 \begin{center}
 \includegraphics[bb=0 0 388 138, scale=.5]{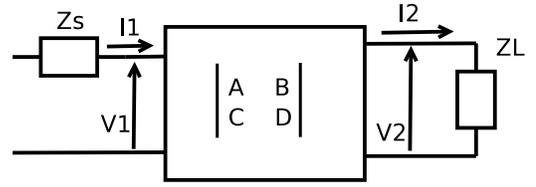}
 % ABCD.eps: 101x0 pixel, 300dpi, 0.86x0.00 cm, bb=0 0 388 138
 \caption{The ABCD parameters for a segment.}
 \label{Fig:ABCD}
\end{center}
\end{figure}

\begin{figure}
\begin{center}
 \includegraphics[scale=.225]{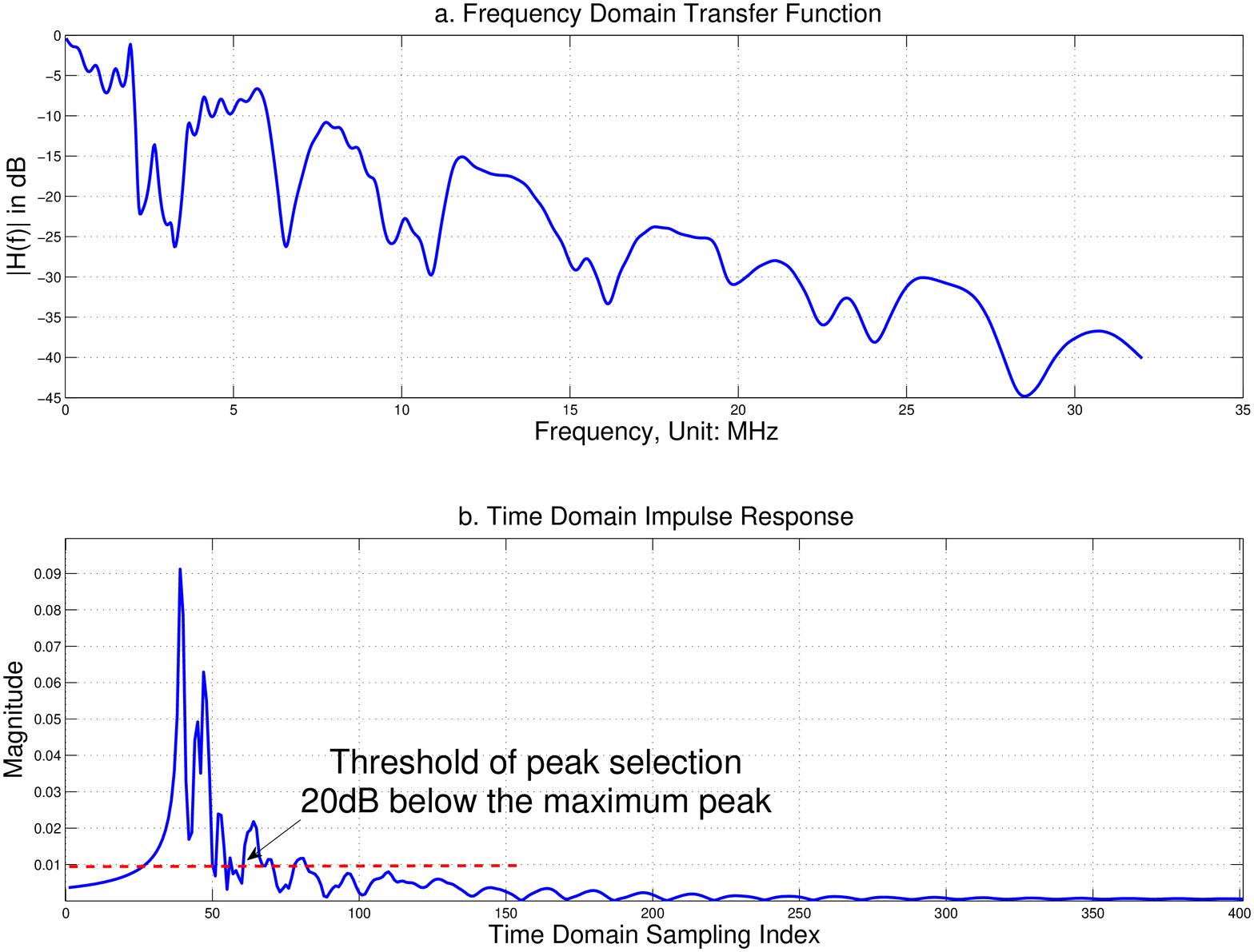}
 % single_channel.eps: 1179668x1179666 pixel, 300dpi, 9987.86x9987.84 cm, bb=   36   148   558   694
 \caption{An example of channel transfer function against the frequency and impulse response against the time domain sampling index, with 100 m transmission distance}
 \label{Fig:single_channel}
\end{center}
\end{figure}

The relations between the inputs and outputs of the 2PN in Fig.~\ref{Fig:ABCD} can be formulated as:
\begin{equation}\label{Equ:ABCD_2PN}
\begin{vmatrix}
V_{1} \\
I_{1} \\
\end{vmatrix}=
\begin{vmatrix}
A & B \\
C & D \\
\end{vmatrix} 
\begin{vmatrix}
V_{2} \\
I_{2} \\
\end{vmatrix}=\textbf{T}_{f}
\begin{vmatrix}
V_{2} \\
I_{2} \\
\end{vmatrix}
\end{equation}
where, $T_{f}$ is called the transmission matrix. The definition of the ABCD parameters can be found in \cite{MTL_Galli} and \cite{MTL_Meng} where it can be seen that \textit{A}, \textit{B}, \textit{C} and \textit{D} are functions of frequency. Then, the transfer function of this segment can be written as:
\begin{equation}\label{Equ:segment_transfer_function}
 H(f)=\frac{Z_{L}}{AZ_{L}+B+CZ_{S}Z_{L}+DZ_{S}}
\end{equation}
The transmission matrix of a shunt segment is:
\begin{equation}\label{Equ:ABCD_shunt}
\textbf{T}_{s}=
\begin{vmatrix}
1 & 0\\
\frac{1}{Z_{in}} & 1\\
\end{vmatrix}
\end{equation} 
where, $Z_{in}=\frac{A}{C}$. 
Thus, the network above can be considered as a series of cascaded segments. After applying the Chain Rule (CR), the transmission matrix for the complete network can be calculated as:
\begin{equation}\label{Equ:Chian_Rule}
 \textbf{T}=\prod_{i=1}^{N} \textbf{T}_{f}^{i}
\end{equation}
where, $\textbf{T}_{f}^{i}$ is the transmission matrix of the $i$th segment. Note that all the shunted segments here have been computed using (\ref{Equ:ABCD_shunt}). By following the above steps, an example transfer function for the frequency domain (\ref{Equ:segment_transfer_function}), and the corresponding time domain impulse response in (\ref{Equ:multipath_prototype}) are shown in Fig. \ref{Fig:single_channel} (a) and \ref{Fig:single_channel} (b) respectively.

The channel transfer function for the above network can be achieved by following the steps in \cite{Multi_path_Zimmermann} and \cite{MTL_Galli}. Based on the modeling result, the following general features of the in-door PLC channel can be seen:
\begin{itemize}
 \item Obvious frequency selective fading in the frequency domain and multipath signal propagation in the time domain.
 \item Higher attenuation at higher frequencies.
 \item Attenuation increasing with the transmitter to receiver distance.
\end{itemize}
These findings are consistent with other literature, e.g. \cite{Mulit_path_Xin}.

\section{Time Domain Channel Modelling Method}
% In this section, a mathematical description of the parameters identified in Section \ref{section: general_feature} are given. Based on the description, the in-door PLC channel as model (\ref{Equ:multipath_prototype}) can be generated. Based on the random topology and the transmission line modelling method in Section \ref{Topology_MTL_model}, a group of network topologies and their corresponding transfer functions and impulse responses can be obtained. The time domain statistical features of the PLC channel are derived from the a large set of generated channels based on randomly generated topologies.
In this section, the methodology of the time domain channel modelling is introduced based on the time domain statistical data. For multipath channels, the number of available paths, the magnitude and delay profile of each path, and the phase of the featured path are often considered in the time domain to form the channel using (\ref{Equ:multipath_prototype}). Based on the random topology and the transmission line modelling method in Section \ref{Topology_MTL_model}, a group of network topologies and their corresponding transfer functions and impulse responses can be obtained. The time domain statistical features of the PLC channel are derived from the a large set of generated channels based on randomly generated topologies. Details about the network topology, cable properties, and extracted channel parameters can be found in \cite{Data_web}.

\subsection{The criterion of selecting a path}
Theoretically speaking there should be infinitely many paths in a single channel impulse response. In order to extract the path features, herein only paths with a magnitude which is larger than a certain threshold (20dB below the maximum peak magnitude, shown as Fig.~\ref{Fig:single_channel} b.) are retained for analysis. This criterion is also used in \cite{Multi_path_Phillipps}. 

\subsection{Channel Cluster}
Regardless of the practical frequency bands used in different PLC standards, a 30 MHz bandwidth which is used in most broadband PLC systems is investigated in this paper. Since discrete time analysis is used in most systems, we use $\tau=3.3\times10^{-8}$s as the sample period to describe the path behaviour in the time domain. In the following parts of this paper, the $k$th sampling point in the time domain is termed as time sample index $k$ and the interval for two adjacent time sampling points is $\tau$. Here, two indicators are investigated: the time sampling index and the magnitude of only path which belongs to the first arrival path.
\begin{figure}
 \begin{center}
 \includegraphics[scale=.225]{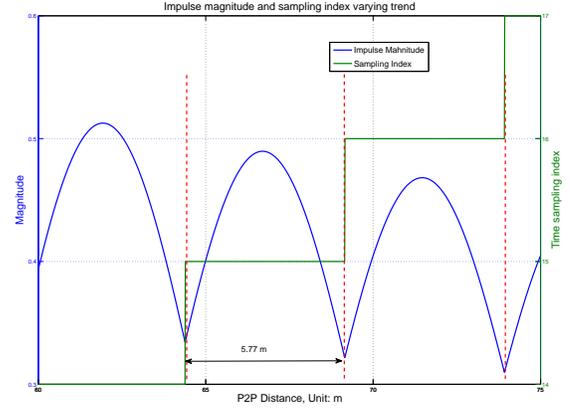}
 % Clusters.eps: 0x0 pixel, 300dpi, 0.00x0.00 cm, bb=   17   216   575   623
 \caption{Varying trends of impulse magnitude and sampling index with P2P distance}
 \label{Fig:Cluster}
\end{center}
\end{figure}
From Fig.~\ref{Fig:Cluster}, we can see that the magnitude of the first arrival path has a obvious periodic decay with increasing P2P distance. The first arrival path sampling index experiences step increase with the increasing P2P distance. The step period is $4.77$ m. Based on this characteristic the concept of a \textit{cluster} is induced herein to indicate the related channel features with transmission distance. Two channels within the same cluster typically present similar features. Thus, channels with transmission distances longer than 10 meters and less then 100 meters are sorted into 20 different clusters, and the cluster index is defined as:
\begin{equation}\label{Equ:cluster_index}
 T_{1}=\lceil\ \left( d -11.92 \right) / 4.77\rceil
\end{equation}
where $d$ is the P2P transmission distance and $\lceil x \rceil$ denotes rounding the elements of $x$ to the nearest integer towards positive infinity.

\subsection{Channel Classification} \label{Channel_Classification}
\begin{figure}
\begin{center}
 \includegraphics[scale=.225]{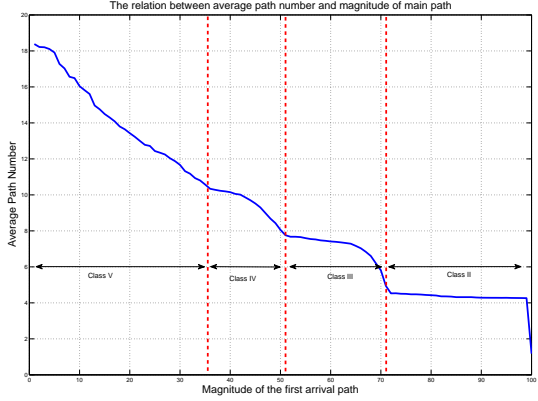}
 % Clusters.eps: 1179666x1310738 pixel, 300dpi, 9987.84x11097.58 cm, bb=   17   216   575   623
 \caption{The step change behaviour of the path number of the $Cluster 10$ channels, the classifying boundaries are given as the red dotted lines}
 \label{Fig:Class}
\end{center}
\end{figure}

As stated in \cite{Statistical_firstpath}, the properties of the first arrival path are important because of its significant power level and the fact that it is usually detectable at the receiver. The reason for this phenomenon is the relatively low frequency-and-distance dependent attenuation and the low number of reflections experienced along the propagation path in comparison with other paths. Thus when observing the magnitudes of the first arrival path (main path) in a cluster, besides the fact that they arrive at the same time sampling index, the magnitudes can be classified into distinct classes using a set of thresholds. There are points of rapid change in the average number of paths as shown in Fig.6.. Thus the channels in a cluster can be classified into 5 distinct classes by using these boundaries as shown by the vertical lines in Fig.6. Futher investigation within a class shows that the average number of paths for channels also shows a step change behaviour. The boundaries for the number of paths are similar to those for the magnitude boundary. For example, the classification method for the $10$th cluster is shown in Fig.~\ref{Fig:Class}. The boundaries for other clusters appear in similar locations.
Note that Class I is not displayed in Fig.~\ref{Fig:Class}, because the channels belonging to Class I are single impulses in the time domain without reflections, which means that there is only one path which is the maximum magnitude channel in Class I. 

\subsection{Distribution of the number of paths}
\begin{figure}
 \begin{center}
 \includegraphics[scale=.225]{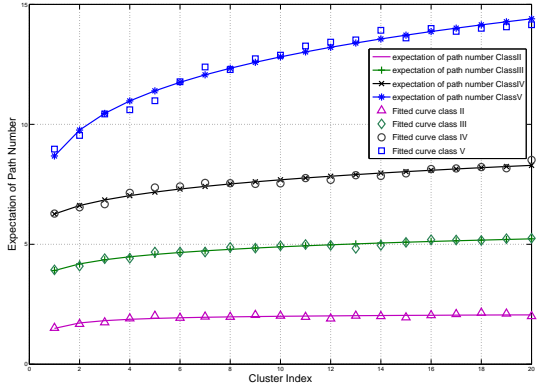}
 % expectation.eps: -415946184x61 pixel, 300dpi, -3521677.50x0.52 cm, bb=   12   230   581   612
 \caption{The mean path number plotted as a function of cluster index which indicates the transmission distance as in equation (\ref{Equ:cluster_index})}
 \label{Fig:expectation}
\end{center}
\end{figure}

\begin{figure}
 \begin{center}
 \includegraphics[scale=.225]{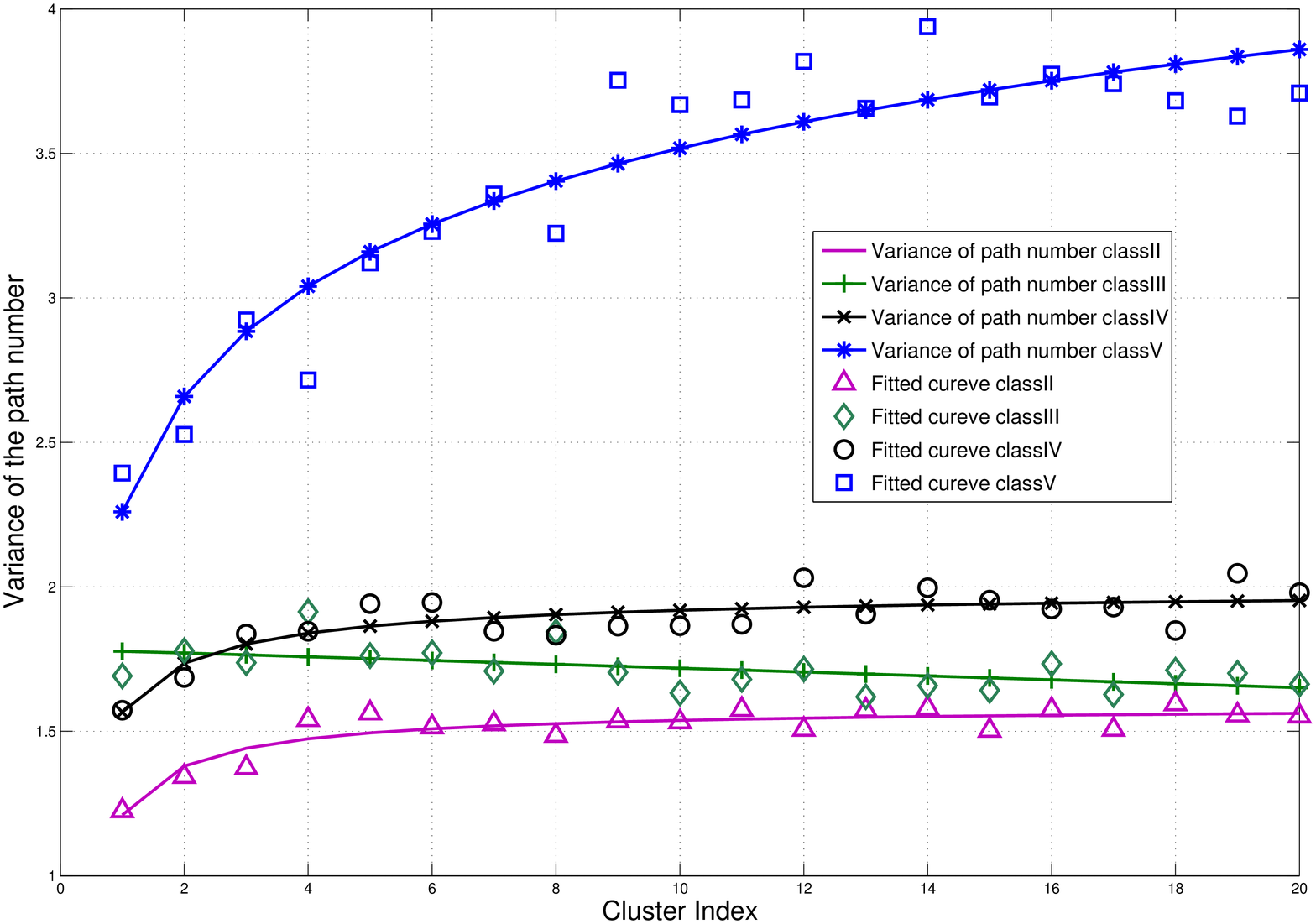}
 % std.eps: 1179666x1179666 pixel, 300dpi, 9987.84x9987.84 cm, bb=   11   205   582   636
 \caption{The variance of the path number plotted as a function of cluster index which indicates the transmission distance}
 \label{Fig:std}
 \end{center}
\end{figure}

The average number of paths is shown in Fig.~\ref{Fig:Class}. But for an individual channel the path number is not fixed to the average value. By observing a group of channels of a certain class (except Class I) in a cluster, the path number for this group shows a Gaussian distribution which can be also seen in other classes of other clusters. Thus, 2 parameter sets, $\mu_{i,k}$ and $\sigma_{i,k}$ ($i=2,3,4,5$), can be used to describe the number of paths for the $i$th class the of $k$th cluster. The trends of these parameters as a function of cluster number (i.e. distance) can be seen in Fig.~\ref{Fig:expectation} and Fig.~\ref{Fig:std}.
Thus, the path number for the $i$th class of the $k$th channel cluster can be written:
\begin{equation}\label{Equ:path_number}
 N_{i,k}=\left[ \mathcal{N} (\mu_{i,k},\sigma^{2}_{i,k}) \right] 
\end{equation}
where $[\cdot]$ means to round towards the nearest integer. In some rare cases that zero or negative numbers will be generated based on (\ref{Equ:path_number}), the number should be dropped and a new positive path number regenerated. In order to obtain the path number for a particular channel by using mathematical approximation, a simple polynomial function is used here to fit the trend of the mean and variance of the path number. The fitted curves can be seen in Fig.~{\ref{Fig:expectation}} and~{\ref{Fig:std}}.  The expression $f(k)=p_{1}k^{p_{2}}+p_{3}$ are use to describe the polynomial function, where $k$ is the cluster index which acts as the argument for the expression, and $p_{1}$, $p_{2}$ and $p_{3}$ are scaling, power and offset parameters respectively. For each class, 2 sets of parameters in \cite{Data_web} are used to describe the expectation $\mu_{i,k}$ and variance $\sigma^{2}_{i,k}$ respectively. 

\subsection{Magnitude features of paths}
The magnitude of the path typically decays as the the time delay increases. In a multipath communications channel, this relationship is called the power delay profile (PDP). In \cite{Statistical_firstpath}, it has been observed that the first arrival path in a PLC channel shows significant differences in magnitude behaviour to the other paths. In this work, the magnitude of the of the first arrival path is investigated separately for each cluster. Then, the PDP of the other paths is given as function of delay spread.

\begin{itemize}
 \item \textbf{Magnitude of First Arrival Path: } The magnitude of the path depends on the how far the signal travels through the network. Thus, the magnitude of the first arrival path will generally decrease as the cluster index increases. This property is shown in Fig.~ \ref{Fig: Mainpath_magnitude_fit}. 
\begin{figure}
 \begin{center}
 \includegraphics[scale=0.225]{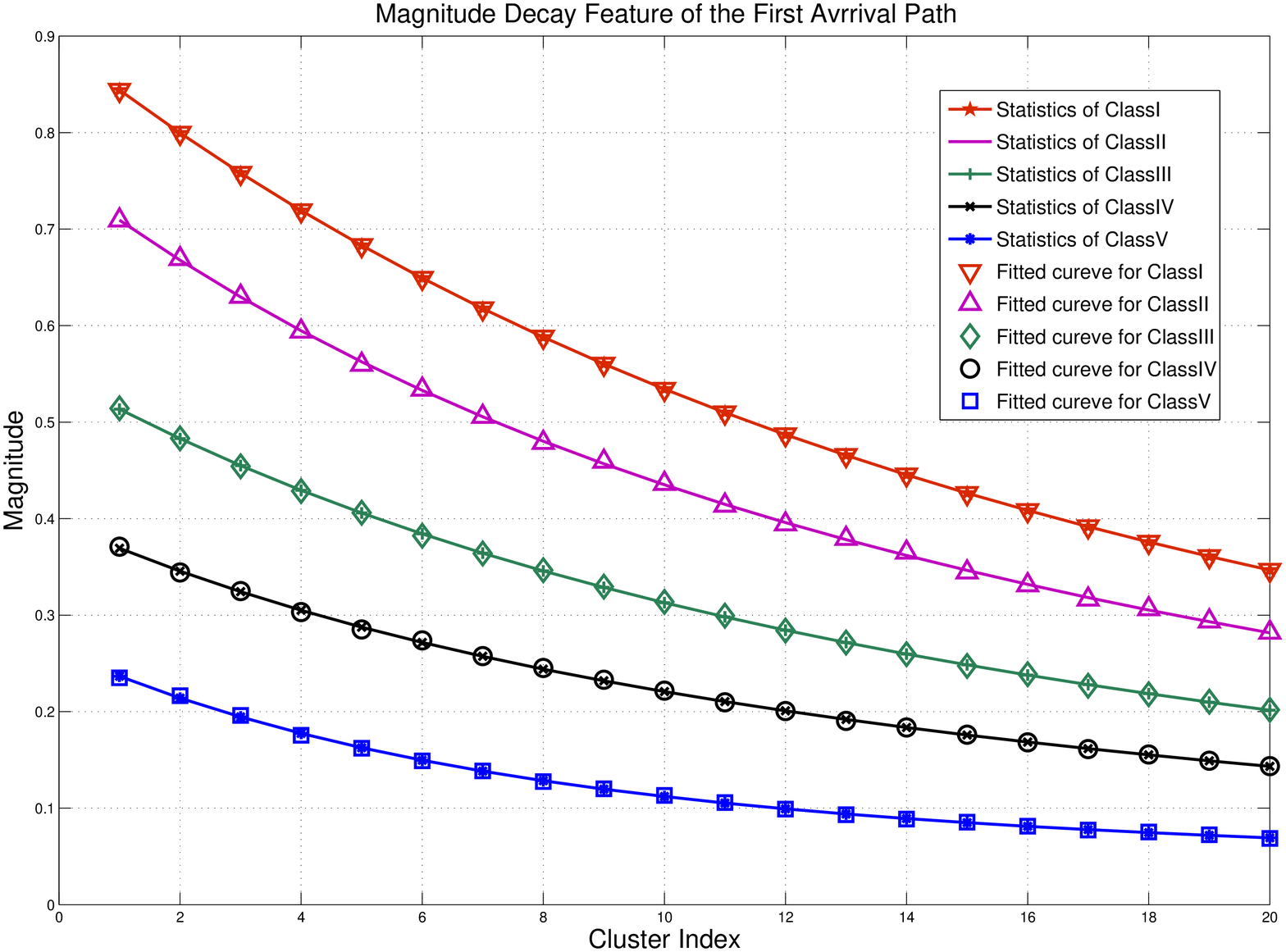}
 % magnitude_class_1.eps: 13215850x15851454 pixel, 300dpi, 111894.20x134208.97 cm, bb=   15   199   579   641
 \caption{Magnitude Decay for the first arrival path as a function of cluster index which indicates the transmission distance}
 \label{Fig: Mainpath_magnitude_fit}
\end{center}
\end{figure}
In Fig.~\ref{Fig: Mainpath_magnitude_fit}, 5 double exponential functions are used to fit the magnitude decay with the cluster index. Therefore, the average magnitude of the $i$th class of $k$th cluster can be expressed as:
\begin{equation} \label{Equ:magnitude_mainpath}
 I_{i,k}=a^{M}_{i}e^{b^{M}_{i}k} + c^{M}_{i}e^{d^{M}_{i}k}
\end{equation}
where the parameters $a^{M}_{i}$, $b^{M}_{i}$, $c^{M}_{i}$ and $d^{M}_{i}$ are given in \cite{Data_web} and $k$ is the cluster index. As noted above, the main path in Class I arrives at the destination without reflections. Thus the first arrival path magnitude for Class I purely depends on the attenuation. For the main path magnitudes of other classes, the power of the transmitted signal may be split by branches at the junctions, and absorbed at branch terminations. Thus, the magnitude for these paths show a random Rayleigh distribution according to \cite{Otherpath_rayleigh}, with the average magnitude in (\ref{Equ:magnitude_mainpath}) as the parameter. It should be understood that the Rayleigh distribution covers multiple network realizations and that the path amplitudes in one network are expected to be relatively stable for long time periods.

 \item \textbf{Power Delay Profile of Other Paths: } Paths experience random reflection and delay in the other classes, thus the magnitudes for these paths do not show step change features. Thus, for these paths, only the magnitude characteristics of the propagation distance (cluster) is investigated. The average magnitudes of different classes within a cluster present very similar decay features, thus, in this paper we do not study the magnitude differences between classes. To simplify the processing herein the time sampling index is used to indicate the propagation distance. The average magnitude decay trends of these paths also can be described by double exponential functions, and Fig.~\ref{Fig: Cluster_Average_fit} shows the fitted double exponential curves for the Cluster 1 and 20 as examples. 
\begin{figure}
 \begin{center}
 \includegraphics[scale=.225]{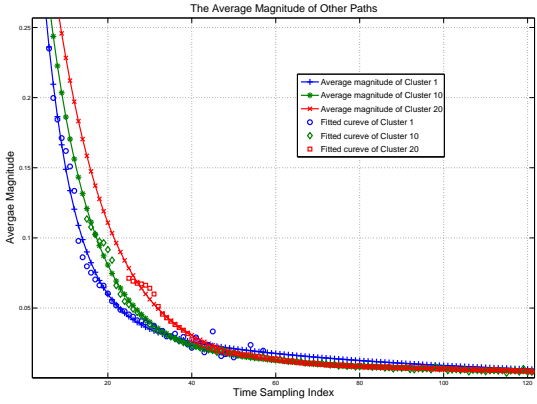}
 % Cluster_Average_fit.eps: 3407920x2097198 pixel, 300dpi, 28853.72x17756.28 cm, bb=   61   259   532   582
 \caption{Average Magnitude VS time delay of Cluster 1, Cluster 10 and Cluster 20}
 \label{Fig: Cluster_Average_fit}
\end{center}
\end{figure}
The double exponential function for these paths is written as:
\begin{equation} \label{Equ:magnitude_otherpath}
 I_{k,j}=a^{O}_{k}e^{b^{O}_{k}j} + c^{O}_{k}e^{d^{O}_{k}j}
\end{equation}
Where $k$ is the cluster index and $j$ is the time sampling index. The values of $a^{O}_{k}$, $b^{O}_{k}$, $c^{O}_{k}$ and $d^{O}_{k}$ are given in \cite{Data_web}.
\end{itemize}
Due to the random signal reflection behaviour through the network, the magnitudes of these paths follow a Rayleigh fading distribution according to \cite{Otherpath_rayleigh}. Thus, the Rayleigh distribution is used to generate the path magnitude with the average magnitude in (\ref{Equ:magnitude_otherpath}) as the parameter. Again, the Rayleigh samples are used to generate different network realizations, the channel for one network is considered to be static with time.

\subsection{Interval Distribution}
\begin{figure}
 \begin{center}
 \includegraphics[scale=.225]{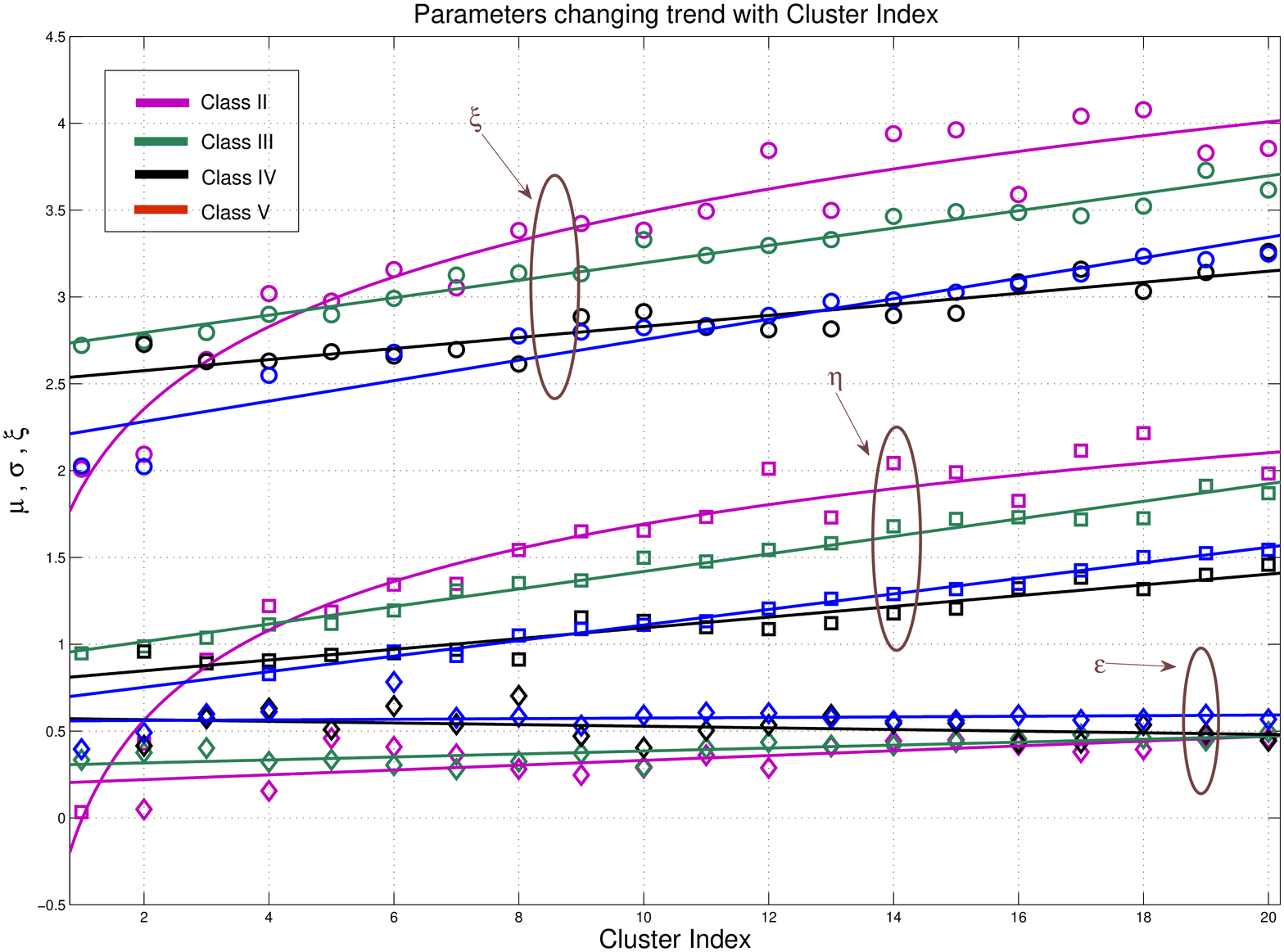}
 % gev_fit.eps: -2022364616x52 pixel, 300dpi, -17122688.00x0.44 cm, bb=  -44   175   641   667
 \caption{The Parameters $\xi$, $\eta$ and $\varepsilon$ for the Time Interval plotted as a function of Cluster Index}
 \label{Fig: GEV_parameters_trend}
\end{center}
\end{figure}

The main cause of the multipath delay interval is due to differences in multipath signal propagation distances. Different signals arrive at different time points which can be indicated by the time sampling index. From the statistical results, it can be found that the path interval distribution follows a Generalized Extreme Value (GEV) distribution. Normally, 3 parameters are used to describe a GEV distribution. Thus the interval PDF of the $i$th class of $k$th cluster can be written as:
\begin{eqnarray} \label{Equ:GEV}
 f_{gev}\left(x;\epsilon_{i,k},\eta_{i,k},\xi_{i,k} \right) \!\!\!\! &=& \!\!\!\! \frac{1}{\eta_{i,k}}\!\!\left(\!\!1+\xi_{i,k}\left(\frac{x-\epsilon_{i,k}}{\eta_{i,k}} \right)\!\!  \right) ^{-\frac{1}{\xi_{i,k}}-1} \nonumber\\
 & & \cdot e^{-\left(1+\xi_{i,k}\left(\frac{x-\epsilon_{i,k}}{\eta_{i,k}} \right)  \right) ^{-\frac{1}{\xi_{i,k}}}}
\end{eqnarray}
How the parameters $\epsilon_{i,k}$, $\eta_{i,k}$ and $\xi_{i,k}$ vary with the cluster index can be seen in Fig.~\ref{Fig: GEV_parameters_trend}.
From Fig.~\ref{Fig: GEV_parameters_trend}, besides the polynomial fitting for $\eta_{2,:}$ and $\xi_{2,:}$ of the Class II, the other parameters are approximated by linear functions of $k$. All the fitted function parameters can be found in \cite{Data_web}. There is no path interval specified for channels of Class I since only one path exists for these channels. 

\subsection{Sampling Rate Conversion}
According to the previous steps, a sampled sequence of paths for a fixed sample rate system can be obtained. As can be seen from the above description of path power delay profile and the path interval, these properties depend on the value of the time domain sampling index. But in practice, different sampling rates may be required e.g. for smart grid systematic operating over narrower bandwidth. Here a method based on the farrow structure filter is given for a user to apply this model to different sampling rates. Assume that the original sampling time is $t_{a}$ and the required sampling time is $t_{a}^{'}$. Thus the conversion rate is $R=t_{a}/t_{a}^{'}$. The structure and designing of the Farrow filter can be found in \cite{farrowsrc}. To simplify the implementation of the sampling rate conversion, the coefficients of the Farrow filter can be obtained by using the function \textbf{mfilt.farrowsrc} in Matlab DSP system toolbox and the conversion rate $R$ is the input argument for function. The channel path sequence under the expecting sampling time $t_{a}^{'}$ can be obtained by passing the original sequence through the Farrow filter.

\subsection{Cable Losses}
According to the previous steps, a sequence of paths can be generated with suitable magnitudes and intervals based on the statistical models above. The sequence can be used to evaluate the impact of multipath fading in a powerline channel. The generated sequence was a perfect impulse shape. But for a multipath channel in practice, it is impossible to approach a perfect impulse due to frequency dependent path-loss. In powerlines, the loss mainly comes from the power dissipation effect brought by the conductor and insulator properties of cable which we call cable losses. In \cite{Multi_path_Zimmermann}, it has been stated that the average cable loss is determined by the real part of the propagation constant and signal propagation distance. Since the propagation constant of the cable is highly frequency dependent, the cable loss in the powerline channel will also be dependent on frequency and distance. According to \cite{Multi_path_Zimmermann}, the cable losses in the frequency domain of a powerline can be approximately written as:
\begin{equation} \label{Equ:cable_losses}
 A\left(f \right)=e^{-\left( a_{0}+a_{1}\cdot f^{k}\right) } e^{-jb_{0}f}
\end{equation}  
where $a_{0}$, $a_{1}$ and $k$ are defined as the attenuation factor. Generally, $a_{0}$ and $a_{1}$ are linear functions of path propagation $d$ for a given cable, and can be calculated by the cable parameters in \cite{Data_web}. The function $A(f,d)$ thus is the propagation attenuation of a certain signal at a given frequency point. Unlike the cable loss in \cite{Multi_path_Zimmermann}, in this paper the phase trend is also given which is determined by cable properties and propagation distance. The linear relation can be described by the parameter $b_{0}$, $b_{0}$ also follows a linear relation with the path propagation $d$. The parameter $b_{0}$ can be found in \cite{Data_web}. Since $a_{0}$, $a_{1}$, $k$ and $b_{0}$ are functions of path propagation distance, the cable loss $A(f)$ will also be a function of the path propagation distance. Based on the parameters in \cite{Data_web}, the cable loss will increase with the propagation distance. Since the arrival time $\tau_{i}$ of $i$th path can be obtained by first arrival path and the path interval, the propagation distance for each path in a certain channel can be calculated by $d=v\tau_{i}$ m. $v$ is the TEM wave propagation speed in the cable, which can be calculated according to the permittivity of the insulator of the cable. Since in this paper, the time domain properties are of interest, the cable loss effect in the time domain can evaluated as the inverse Fourier transform of (\ref{Equ:cable_losses}), written as $A_{t}\left(t,v\tau_{i} \right) $. Thus, the time domain channel impulse response with cable losses is:
\begin{equation} \label{Equ:CIR_cable_losses}
 h(t)=\sum_{i=1}^{N} I_{i}\cdot \left[A_{t}\left(t,vT_{i} \right)\otimes\delta(t-T_{i}) \right]
\end{equation}  
Using equation (\ref{Equ:CIR_cable_losses}), $h(t)$ can fully present the multipath fading, cable losses, and phase of the featured paths.  

%%%%%%%%%%%%%%%%%%%%%%%%%%%%%%%%%%%%%%%%%%%%%%%%%%%%%%%%%%%%%%%%%%%%%%%%%%%%%%%%%%%%%%%%%%%%
%   Simulation Part
%%%%%%%%%%%%%%%%%%%%%%%%%%%%%%%%%%%%%%%%%%%%%%%%%%%%%%%%%%%%%%%%%%%%%%%%%%%%%%%%%%%%%%%%%%%%
\section{Application for Communication Theory Capacity Evaluation}
As mentioned in \cite{Statistical_firstpath}, the communications technology can not be built for specific network case. However, based on the statistical model proposed in our paper, the PLC network can be evaluated using high level attributes which average over the details of the network topology and the configuration information. To prove the suitability of the proposed model for PLC research, modelling results with the both statistical model and traditional transmission line theory using random topologies are compared in the frequency domain. To further verify the statistical model, the average Shannon capacities and capacity distributions are derived and compared with results that are based on transmission line theory. 

\subsection{General comparison of both frequency domain and time domain response}
\begin{figure}
\begin{center}
 \includegraphics[scale=0.225]{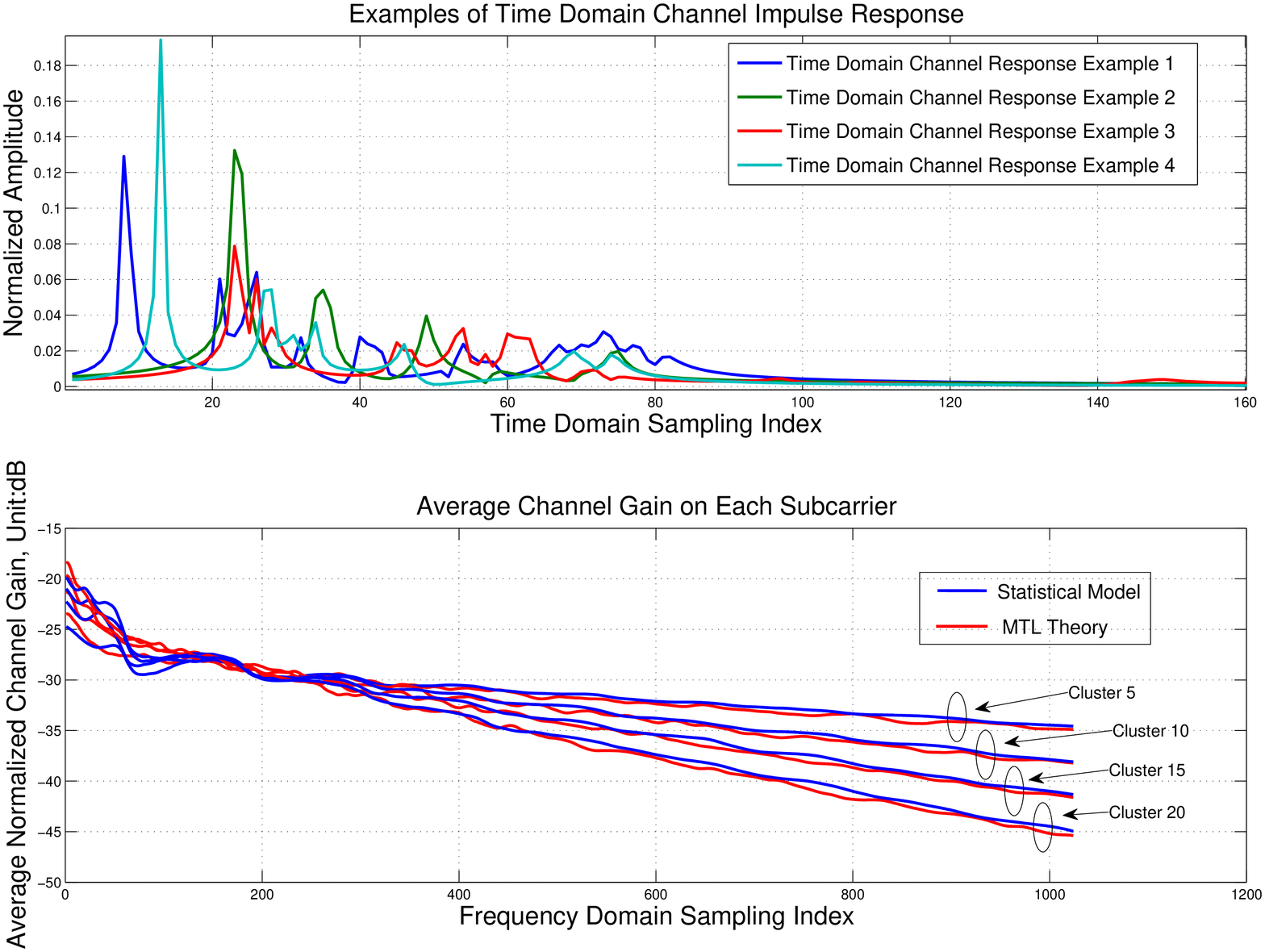}
 % Cluster_Compare.eps: 553670640x0 pixel, 300dpi, 4687745.00x0.00 cm, bb=  -27   128   622   713
 \caption{Statistical Model Verification for Cluster 5, 10, 15 and 20}
 \label{Fig: cluster_10_20_30}
\end{center}
\end{figure}

In Fig.~\ref{Fig: cluster_10_20_30} (a), three time domain channel impulse response examples based on the statistical model are given. The Channel Impulse Responses (CIR) show clear multipath features and a general decrease trend of path magnitude with time delay that fits the features described in Section III. In Fig.~\ref{Fig: cluster_10_20_30} (b), the average channel gains based on the statistical model and transmission line theory are compared. The results are based on the power cables and network topology introduced in Section \ref{section: general_feature}.

The average channel gains in \ref{Fig: cluster_10_20_30} Fig. (b) are derived from the normalized transfer function in the frequency domain. The average channel gain of clusters 5, 10, 15 and 20 show obvious attenuation with increasing frequency. Also the cable loss increases with the cluster index which indicates the propagation distance of the channel. The trends presented from the statistical modelling show good consistency with the results from transmission line theory.

\subsection{Capacity comparison}
In order to test the suitability of the statistical model in PLC research, the average capacities and CDFs of channels capacity for different clusters have been simulated. To make sure the comparison is meaningful, here the same transmission power according to FCC part 15 \cite{Sony_capacity} is used and the in office noise scenario introduced in \cite{Dirk_noise} is also included. 
\begin{figure}
 \begin{center}
 \includegraphics[scale=.325]{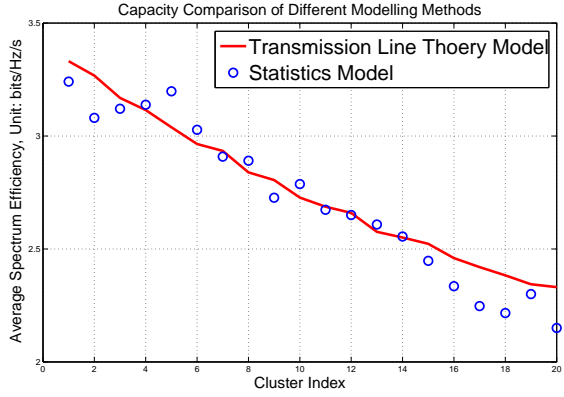}
 % capacity_cluster.eps: -415946184x61 pixel, 300dpi, -3521677.50x0.52 cm, bb=   60   242   534   599
 \caption{The capacity trend as a function of cluster index which indicates the the transmission distance}
 \label{Fig: capacity_cluster}
\end{center}
\end{figure}

\begin{figure}
 \begin{center}
 \includegraphics[scale=.225]{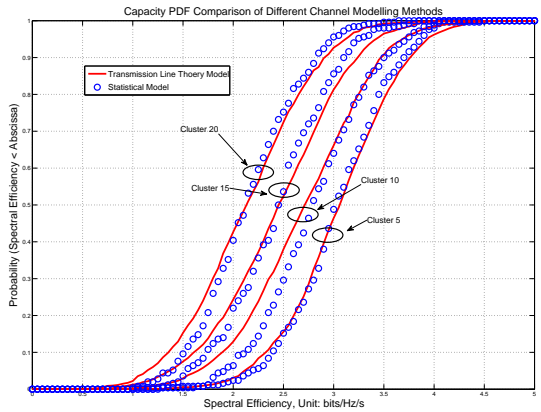}
 % capacity_cluster.eps: -415946184x61 pixel, 300dpi, -3521677.50x0.52 cm, bb=   60   242   534   599
 \caption{CDFs of the capacities based on the statistics model and transmission line theory}
 \label{Fig: CDF_Cluster}
\end{center}
\end{figure}

From results in Fig.~\ref{Fig: capacity_cluster}, the average Shannon capacities based on the statistical model and transmission line theory are shown. As can be seen from the figure, the transmission capacities from both models decreases with increasing transmission distance, with only sight differences in the results. In Fig.~\ref{Fig: CDF_Cluster}, the CDF of the capacity distributions are given. From the CDF we can see that the statistical model is also capable of presenting the differences of channel conditions in the PLC network, not just the average trend. Also the comparison between results from statistical and transmission line theory models show that the proposed model works properly in presenting the range of different channel conditions. 

%%%%%%%%%%%%%%%%%%%%%%%%%%%%%%%%%%%%%%%%%%%%%%%%%%%%%%%%%%%%%%%%%%%%%%%%%%%%%%%%%%%%%%%%%%%%
%
%%%%%%%%%%%%%%%%%%%%%%%%%%%%%%%%%%%%%%%%%%%%%%%%%%%%%%%%%%%%%%%%%%%%%%%%%%%%%%%%%%%%%%%%%%%%
\section{Application for Impulse Noise Modelling}
\begin{figure}
 \begin{center}
 \includegraphics[scale=.3]{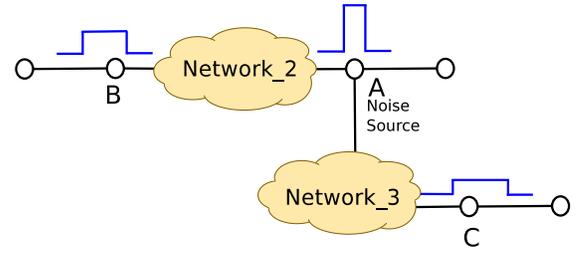}
 % Impulse_spread.eps: -415946184x61 pixel, 300dpi, -3521677.50x0.52 cm, bb=0 0 698 315
 \caption{The Impulse Noise Spread Model in PLC Network}
 \label{Fig: impulse_noise_spread}
\end{center}
\end{figure}

Impulse noise in PLC is composed of periodic and aperiodic impulsive noises. The periodic impulsive noise which is caused by power converters in power supplies and by rectifiers operatings in the alternating voltage current network. Sources for aperiodic impulsive noise are switched power supplies, the turning on/off of appliances, and so on. In \cite{Impulse_noise_Zimmermann} a Markov chain is used to simulate the impulse noise behaviour. Based on \cite{Impulse_noise_Zimmermann} the impulse noise will appear independently at each PLC node and socket in general. But considering the practical situation in PLC network, the impulse noise will propagate in a given PLC network through the power cables. Thus, the impulse noise on different sockets or nodes will be correlated. As shown in Fig.~\ref{Fig: impulse_noise_spread} impulse arises at Socket A due to the switch on/off of this socket. After spreading to Socket B and Socket C through Network\_2 and Network\_3, the impulse magnitude will be reduced, while the delay spread will increase. To build a more realistic signal propagation environment in PLC, a correlated impulse noise model is thus necessary. This is particularly important for concepts such as the relay enhanced PLC network as described in \cite{ICC_BoTan}. Assume Socket B and Socket C are the transceivers and Socket A is the relay node for the bi-directional data transmission protocols in \cite{ICC_BoTan}. In the bi-directional scenario, both data transmission and reception will be disturbed when the impulse noise arises, since the impulse noise will propagate simultaneously with the signal. Thus, the model of impulse noise is particularly important for capacity evaluation of bi-directional relay protocols in PLC networks. Also, with a more realistic impulse noise, high performance noise cancellation schemes could be developed which exploit knowledge of the correlation of the impulse noise.

\begin{figure}
 \begin{center}
  \includegraphics[scale=.225]{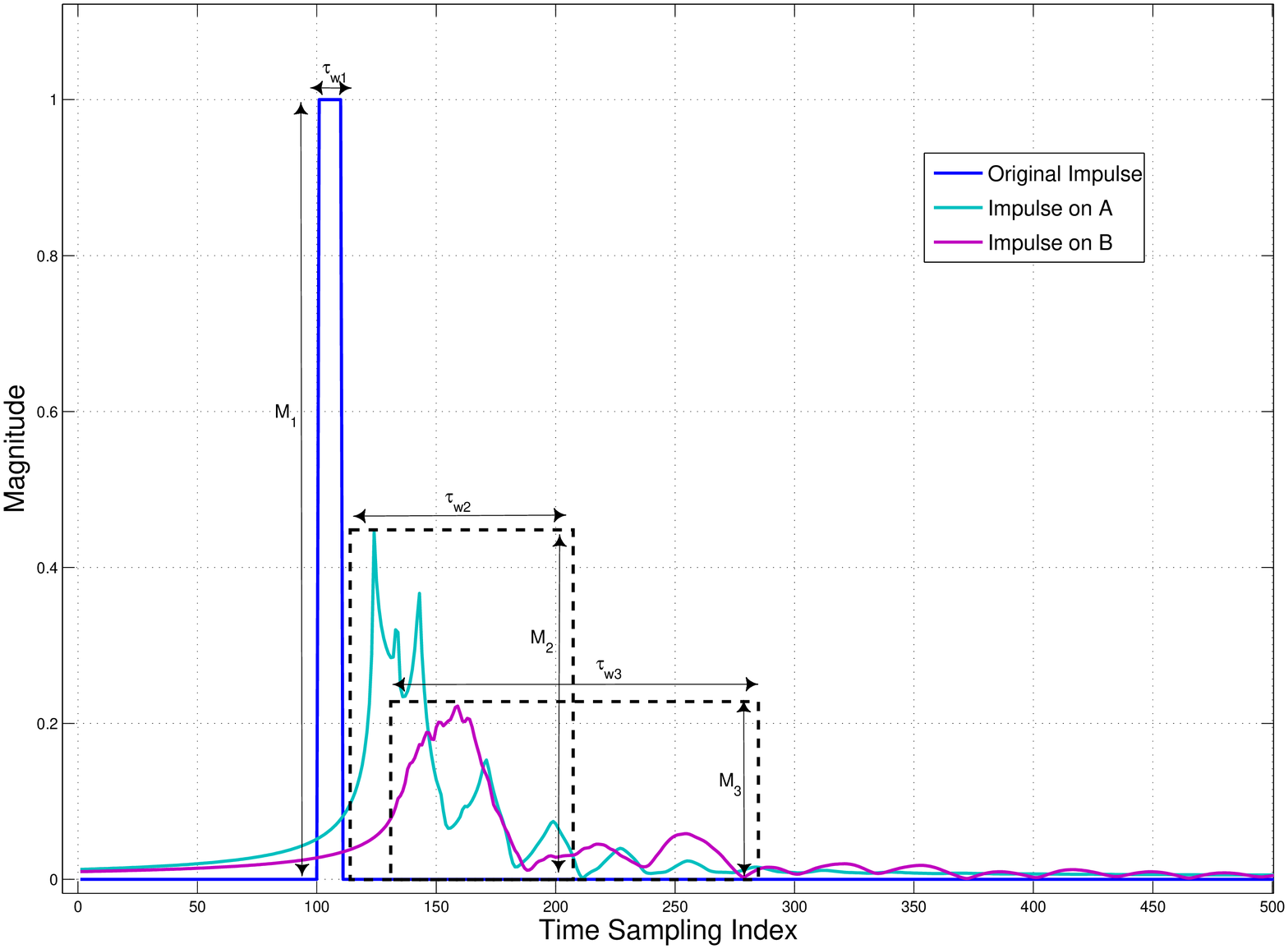}
  % Impulse_spread.eps: -415946184x61 pixel, 300dpi, -3521677.50x0.52 cm, bb=0 0 698 315
  \caption{Example of Impulse Noise Generation in PLC Network}
  \label{Fig: impulse_noise_Cluster23}
 \end{center}
\end{figure}

Considering a realistic scenario, if we intend to obtain the exact impulse noise on each socket channel knowledge between every socket pair should be known, which is complex to evaluate in practice. If we apply the time domain statistical model of the PLC network into impulse noise modeling, the only knowledge required is the distance profile between each socket pair. The impact of the network topology and impedance on the terminals are integrated into the statistical data. Assume Network\_2 and Network\_3 in Fig.~\ref{Fig: impulse_noise_spread} are belong to Cluster 10 and Cluster 20 respectively. Here, we assume the the rectangular impulse originates at node A. Two channels are generated according to the proposed statistical model for Cluster 10 and Cluster 20 respectively. After the convolution operation, the magnitudes and time spreads of an impulse on node B and C are shown as Fig.~\ref{Fig: impulse_noise_Cluster23}. Unlike current impulse noise modelling methods, the impulse noise at different PLC transmission nodes will be correlated. The simulation result in Fig.~\ref{Fig: impulse_noise_Cluster23} shows that the amplitude of the impulse noise is attenuated when it passes through the channel, meanwhile the time span become wider. 
The details of the attenuated amplitude $M_{i}$ and spread time $\tau_{wi}$ can be calculated by using the proposed statistical channel model. Considering amplitude attenuation and time spread of the correlated impulse noise in PLC networks, related technologies such as system capacity
calculation, channel coding and modulation can be reconsidered in PLC.

%%%%%%%%%%%%%%%%%%%%%%%%%%%%%%%%%%%%%%%%%%%%%%%%%%%%%%%%%%%%%%%%%%%%%%%%%%%%%%%%%%%%%%%%%%%%
%
%%%%%%%%%%%%%%%%%%%%%%%%%%%%%%%%%%%%%%%%%%%%%%%%%%%%%%%%%%%%%%%%%%%%%%%%%%%%%%%%%%%%%%%%%%%%
\section{Conclusion and Future Work}
In this paper, a statistical channel modelling method in the time domain is proposed for PLC networks. First, the channels are sorted into different categories based on the P2P distance and the magnitude of the first arrival path. Second, the multipath parameters such as path magnitude and path interval are extracted to build the time domain impulse response. Furthermore, a Farrow structure filter is proposed to make sampling rate conversion according to the practical scenario requirements. Finally, the proposed statistical model is used for communications theory evaluations and correlative impulse noise modelling in PLC networks. The proposed statistics integrate the impact of topology and terminal loads. By comparing with the modelling results from the transmission line method, the statistical model is proved to accurately capture the path delay and the average attenuation. Also, the Shanon capacity derived from statistical and transmission line models shows that the proposed model is a feasible tool for PLC research. Based on this statistical model, fast and efficient studies of deployment, coverage and capacity of the PLC network can be carried out. In order to extend the proposed methodology to a more general application scenarios, especially for the Smart Grid, more channel magnitude and time spread types for different power grids should be considered in future where the main issue is the impact of different branch densities on path magnitudes and path intervals.

\ifCLASSOPTIONcaptionsoff
  \newpage
\fi

\end{document}